\documentclass[10pt,sigconf]{acmart}

\makeatletter
\def\@ACM@checkaffil{% Only warnings
    \if@ACM@instpresent\else
    \ClassWarningNoLine{\@classname}{No institution present for an affiliation}%
    \fi
    \if@ACM@citypresent\else
    \ClassWarningNoLine{\@classname}{No city present for an affiliation}%
    \fi
    \if@ACM@countrypresent\else
        \ClassWarningNoLine{\@classname}{No country present for an affiliation}%
    \fi
}
\makeatother

\setcopyright{none}
\author{Saurab Dulal}
\orcid{0000-0003-2457-1214}
\affiliation{%
  \institution{University of Memphis}
}
\email{sdulal@memphis.edu}

\author{Tianyuan Yu}
\affiliation{%
  \institution{UCLA}
}
\email{tianyuan@cs.ucla.edu}

\author{Siqi Liu}
\affiliation{%
  \institution{UCLA}
}
\email{tylerliu@g.ucla.edu}

\author{Adam Robert Thieme}
\affiliation{%
  \institution{University of Memphis}
}
\email{athieme@memphis.edu}

\author{Lixia Zhang}
\affiliation{%
  \institution{UCLA}
}
\email{lixia@cs.ucla.edu}

\author{Lan Wang}
\affiliation{%
  \institution{University of Memphis}
}
\email{lanwang@memphis.edu}
\usepackage[hyphenbreaks]{breakurl}
\usepackage{stfloats}
\usepackage{subfigure}
\usepackage{listings}
\usepackage{multirow,tabularx}
\usepackage{graphicx}
\usepackage{algorithmic}
\usepackage{graphicx}
\usepackage{textcomp}
\usepackage{xcolor}
\usepackage{tabularx}
\usepackage{tabularx,booktabs,ragged2e}
\usepackage{listings}
\usepackage{makecell}
\usepackage{url}
\usepackage{listings}
\usepackage{amsmath}
\usepackage{comment}
\usepackage{tikz}
\usepackage[inline]{enumitem}

\usepackage{hyperref}
\usepackage{xcolor}
\usepackage{etoolbox}
\usepackage{xstring}
\DeclareListParser{\doslashlist}{/}
\newcounter{ndnNameComponentCounter}%
\newcommand{\name}[1]{{%
  \setcounter{ndnNameComponentCounter}{0}%
  \renewcommand{\do}[1]{{%
    \ifnumgreater{\value{ndnNameComponentCounter}}{0}{\allowbreak/}{}%
    \ifnumodd{\value{ndnNameComponentCounter}}{}{}%
    \detokenize{##1}}%
    \stepcounter{ndnNameComponentCounter}}%
``{\fontfamily{cmtt}\small\selectfont\IfBeginWith{#1}{/}{/}{}\doslashlist{#1}}''%
}}

\usepackage{ifthen}
\newboolean{showcomments}
\setboolean{showcomments}{true}
\ifthenelse{\boolean{showcomments}}
{ \newcommand{\mynote}[3]{
    \protect\fbox{\bfseries\sffamily\scriptsize#1}
    {\small$\blacktriangleright$\textsf{\emph{\color{#3}{#2}}}$\blacktriangleleft$}}}
{ \newcommand{\mynote}[3]{}}

\newcommand{\sd}[1]{\mynote{Saurab}{#1}{red}}

\begin{document}
\include{macros}

%% allowing the author to define a "short title" in page headers.
\title{Enhancing NAC-ABE to Support Access Control for mHealth Applications and Beyond}

%\begin{CCSXML}
%<ccs2012>
%  <concept_id>10010520.10010553.10010562</concept_id>
%  <concept_desc>Computer systems organization~Embedded systems</concept_desc>
%  <concept_significance>500</concept_significance>
%  <concept_id>10010520.10010575.10010755</concept_id>
%  <concept_desc>Computer systems organization~Redundancy</concept_desc>
%  <concept_significance>300</concept_significance>
%  <concept_id>10010520.10010553.10010554</concept_id>
%  <concept_desc>Computer systems organization~Robotics</concept_desc>
%  <concept_significance>100</concept_significance>
%  <concept_id>10003033.10003083.10003095</concept_id>
%  <concept_desc>Networks~Network reliability</concept_desc>
%  <concept_significance>100</concept_significance>
%</ccs2012>
%\end{CCSXML}


%\keywords{Information Centric Networking, Named Data Networking, Service Discovery, sync}

\begin{abstract}
    Name-based access control (NAC) over NDN provides fine-grained data confidentiality and access control by encrypting and signing data at the time of data production. 
    NAC utilizes specially-crafted naming conventions to define and enforce access control policies.
    NAC-ABE, an extension to NAC, uses an attribute-based encryption (ABE) scheme to support access control with improved scalability and flexibility. 
    However, existing NAC-ABE libraries are based on ciphertext-policy ABE (CP-ABE), which requires knowledge of the access policy when encrypting data packets. 
    In some applications, including mHealth, the data access policy is unknown at the time of data generation, while data attributes and properties are known. 
    In this paper, we present an extension to the existing NDN-ABE library which can be used by mHealth and other applications to enforce fine-granularity access control in data sharing.
    We also discuss the challenges we encountered during the application deployment, and remaining open issues together with potential solution directions.
\end{abstract}


\maketitle

\pagestyle{plain}

\section{Introduction}
\label{introduction}
Named Data Networking (NDN)~\cite{NDN-TR1, zhang2014named, zhang2018overview} is a new network architecture design which lets applications retrieve desired data by names instead of IP addresses. NDN introduces a data-centric security model that ensures end-to-end security; it also makes data dissemination efficient and robust. 

NDN's data-centric security model enables fine-grained access control through Name-based access control (NAC)~\cite{yu2015name, zhang2018nac}, which encrypts and signs data at its production time. 
NAC uses a combination of asymmetric and symmetric encryption techniques to ensure data confidentiality. It also automates the distribution of data decryption keys by utilizing NDN naming conventions. 
NAC-ABE~\cite{zhang2018nac} is an extension to NAC that incorporates attribute-based encryption (ABE) \cite{goyal2006attribute} scheme to improve its scalability and flexibility \cite{belguith2018phoabe}. ABE schemes offer more precise control over data accessibility than NAC alone by supporting granular attribute-based policies.

However, the existing NAC-ABE library~\cite{zhang2020nacabe} is based on Ciphertext-Policy ABE (CP-ABE)~\cite{bethencourt2007ciphertext}, its data encryption operation requires the knowledge of the access policy.
%%, resulting in significant limitations.
% In certain applications such as mHealth, data attributes and properties are known at data production time, but the access policy, which governs data accessibility, is unknown. Furthermore, two other issues in the library are also discovered: \textbf{i) the absence of packet validation}, which introduces potential risks of unauthorized data production. For example, the Attribute Authority (AA) could produce and distribute an unauthorized decryption key, which would introduce a significant security risk. 
In certain applications such as mobile health (mHealth) \cite{sim2019mobile}, data's attributes and properties are known at production time, but the access policy may be unknown. Furthermore, several other identified issues in the library also impact its usage: 
\textit{i) the absence of packet validation}, which introduces potential risks of unauthorized data production. For example, the Attribute Authority (AA) could produce and distribute an unauthorized decryption key, which introduces a significant security risk;
\textit{ii) the high frequency of content key (CK) generation} per packet which can introduce significant overhead for systems, such as mHealth, that produce sensing data at a high rate; and
\textit{iii) the inability of handling large Decryption Keys (DKEYs) and CKs}, when their packet size exceeds NDN's Maximum Segment Size (MSS). 

To provide fine-grained data access control for mHealth app, we developed solutions to the aforementioned issues. Our contributions can be summarized as follows:
First, we added support for KP-ABE (Key-Policy Attribute-Based Encryption) \cite{goyal2006attribute} to the existing NAC-ABE library. KP-ABE removes the need for knowing the access policy during encryption, thereby avoiding Content Key re-encryption when changing the access control policy. 
Second, we made changes to the naming scheme to incorporate KP-ABE, key segmentation, and other necessary modifications. 
Third, we designed and implemented a CK-Caching producer that reuses CKs wherever applicable and enables applications to configure the periodicity of CK updates based on their specific needs. 
In addition, we also introduced data validation to ensure the authenticity of each packet exchanged by NAC-ABE, and implemented segmentation to handle large DKEYs and CKs when their size exceeds the NDN MSS. 
Furthermore, we identified and discussed several encountered issues to guide future developments in NAC-ABE libraries.

The remaining sections of this paper are structured as follows.
Section \ref{background} provides background information regarding NAC and NAC-ABE. 
In Section \ref{sec:use-case-mhealth}, we present a use-case scenario in the context of mHealth, outlining its goals, assumptions, and security requirements. Additionally, we highlight various issues that we identified within the existing NAC-ABE. 
We explain our design approach and the components involved, including implementation details, in Section \ref{design}. 
Section \ref{sec:discussion-and-lessons-learned} focuses on the discussion of remaining issues and potential future directions for how to solve them. 
Finally, we conclude our work in Section \ref{conclusion-future-work}. 

% % first person rewording
% In Section \ref{background}, we provide background information regarding NAC and NAC-ABE. 
% In Section \ref{sec:use-case-mhealth}, we present a use-case scenario in the context of mHealth, outlining its goals, assumptions, and security requirements. Additionally, we highlight various issues that we identified within the existing NAC-ABE. 
% We explain our design approach and the components involved, including implementation details, in Section \ref{design}. 
% In Section \ref{sec:discussion-and-lessons-learned} we focus on the discussion of remaining issues and potential future directions. 
% Finally, we conclude our work in Section \ref{conclusion-future-work}. 

% % third(-ish) person rewording
% Section \ref{background} provides background information regarding NAC and NAC-ABE. 
% Section \ref{sec:use-case-mhealth} presents a use-case scenario in the context of mHealth, outlining its goals, assumptions, and security requirements. Additionally, it highlights various issues we identified within the existing NAC-ABE. 
% Our design approach and the components involved, including implementation details, are explained in Section \ref{design}. 
% Section \ref{sec:discussion-and-lessons-learned} focuses on the discussion of remaining issues and potential future directions. 
% Finally, Section \ref{conclusion-future-work} concludes our work.

% \lz{given no paper outline, please add the section number where each of the above contributions is described}
\section{Background} 
\label{background}

In this section, we begin by introducing NAC (Name-based Access Control)~\cite{yu2015name, zhang2018nac}, which utilizes NDN naming conventions to automate key management and enable fine-grained access control. Next, we provide an overview of NAC-ABE~\cite{zhang2018nac}, which achieves the design goals of NAC using Attribute-Based Encryption.
%The design of NAC assumes that proper trust relationships among entities in the system have already been established.
%To be more specific, 
%\begin{enumerate*} [label=(\roman*)]
%\item each entity in the system has its own public/private key pair, 
%and 
%\item each entity is able to authenticate the Data packets produced by others through digital signature validation.
%\end{enumerate*}

\subsection{Name-based Access Control (NAC)}
\label{sec:nac-background}

\subsubsection{Design Goals} 
The NAC design aims to achieve the following goals:
\begin{enumerate}[leftmargin=5mm]
\item control data access with fine granularities;
\item automate the data access control process as much as possible;
\item make the system robust against intermittent network connectivity.
\end{enumerate}
NAC achieves the above goals by using NDN's structured, semantically-meaningful naming to express the access control policy and granularity, combined with the used of symmetric and asymmetric cryptographic keys (see Fig.~\ref{fig:nac-model}) 

\subsubsection{Design Overview}

\begin{figure}[tbp]
	\centering
	\includegraphics[scale=0.4]{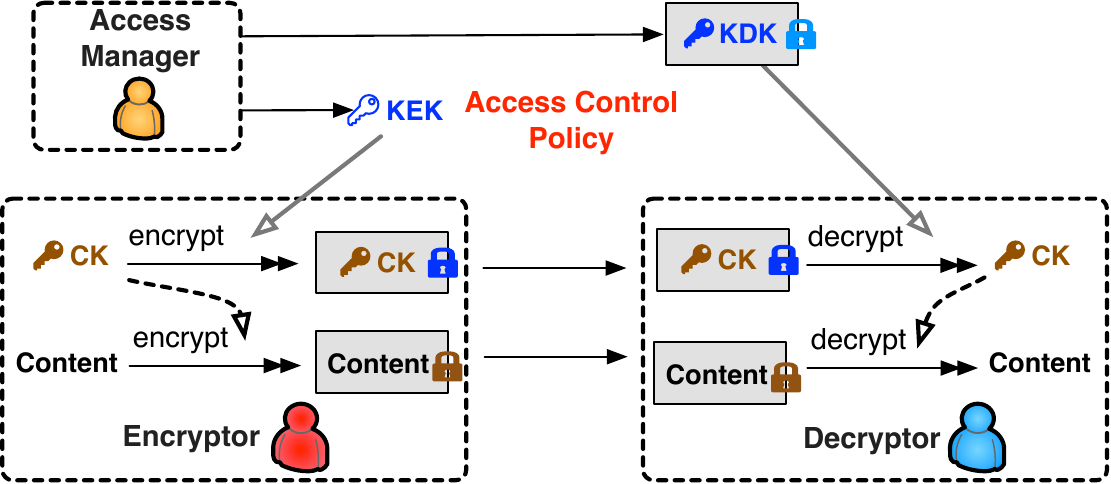}
	\caption{Content encryption and decryption in NAC Scheme 
 % \lz{need to add more explanation}
 }
	\label{fig:nac-model}
\end{figure}

NAC assumes the existence of an \emph{access control manager}\footnote{A data producer can serve as its own access control manager.}, $M_{actrl}$, which defines the access control policies in a given system (see Figure~\ref{fig:nac-model}).
Assuming data is encrypted by its producers, $M_{actrl}$ should provide authorized data consumers with the right decryption key automatically, i.e. with no user involvement.
NAC solved this problem in the following way. For a collection of data that requires access control, the access manager $M_{actrl}$ publish a list of named public and private key pairs, called KEK (key-encryption key) and KDK (key-decryption key), respectively, as NDN data packets.
A KEK's name indicates the \emph{granularity} of the Data packets that can be access controlled by this KEK:
\begin{center}
	KEK Name = \name{/<access-manager-prefix>/NAC/<granularity>/KEK/<key-id>}
\end{center}
%This named policy can be configured or inferred from configuration and data name (see Section~\ref{sec:auto} for an example of it).
A data producer retrieves the KEKs by names, which will be used to encrypt a symmetric key, CK (content key), used to encrypt its data.
Following the NAC naming convention, the encrypted CK data packet is named as follows:
%As a result, an encryption (or decryption) key chain can be established from an producer to a consumer under the control of the access manager.
\begin{center}
	CK Data Name = \name{/<producer-prefix>/CK/<key-id>/ENCRYPTED-BY/<KEK Name>}
\end{center}
On the other hand, the data packets carrying KDKs (key-decryption key) are encrypted using the public key of each authorized data consumer. The name of KDK data packet also follows a well defined naming convention:
\begin{center}
	KDK Data Name = \name{/<access-manager-prefix>/NAC/<granularity>/KDK/<key-id>/ENCRYPTED-BY/<consumer-public-key>}.
\end{center}
%In NAC, the access manager, producer, and consumer are different in terms of their function.
%In practice, they can be a single entity based on specific application scenarios.
%For instance, in our battlefield scenario, the command center is the access manager, producer (when sending commands to others), and consumer (for received responses).
Following the naming convention, an authorized consumer can construct the KDK data name to fetch it, uses its own private key to decrypt to obtain the KDK, then uses the KDK to get the CK to decrypt the access controlled data.

\subsubsection{Properties}
NAC's fine-grained access control requires that the access control manager $M_{actrl}$ generate KEK/KDK pairs with the granularity of the access controlled data.
The finer data granularity the control policy specifies, the more KEK/KDK pairs must be generated, each controlling access to a specific amount of data.
%% The more specific the KEK name is, the fewer Data packets can be decrypted using the corresponding KDK.
% The more specific the KEK name, the smaller the set of Data packets which can be decrypted by the corresponding KDK.
$M_{actrl}$ distributes the data decryption keys by publishing them in encrypted form, using the public key of each authorized consumer. 
Therefore, the number of KDKs $M_{actrl}$ has to publish goes up linearly with the number of authorized consumers.
%%NAC also allows the access manager to control the access through KDK distribution.
%%The more KDKs a consumer has, the more Data packets it can access.

NAC automates access control by encrypting and then publishing data decryption keys as normal NDN Data packets which can only be decrypted by authorized data consumers, and by establishing naming conventions to enable authorized data consumers to systematically construct the names for fetching KDK packets.
%%As long as the named KEKs have been published in the network, producers can automatically retrieve them by name to encrypt data and CKs properly.
%The access manager and producers also publish the KDKs and CKs respectively, so that consumers can follow the key names encoded in the encrypted Data packets to retrieve decryption keys and construct the decryption key chain automatically.
Since all the keys are published as regular NDN Data packets, NDN's data-centric security enables access control managers to freely distribute these Data packets over untrusted networks, or stored in untrusted data repositories.
Fetching data by names ensures that, as long as any copy for those keys exist in the network, they can always be retrieved even with intermittent connectivity.

\subsection{NAC-ABE}
\label{sec:nacabe-background}

The NAC design uses traditional public-key cryptography (e.g., RSA) \cite{diffie2022new} to generate KEK/KDK keys.  To achieve better scalability and flexibility, Zhang et. al. \cite{zhang2018nac} proposed the use of Ciphertext-Policy Attribute-Based Encryption (CP-ABE)~\cite{bethencourt2007ciphertext} to support data access control. 
In this section, we give an overview of this NAC-ABE implementation~\cite{zhang2020nacabe}, which is referred to as \emph{NAC-ABE-2020} in this paper.

Before delving into the details, we first provide a brief background on ABE (Attribute-Based Encryption) and CP-ABE.  
ABE is a public-key encryption technique~\cite{sahai2005fuzzy}.  \textbf{Attributes} are descriptive strings, such as professor, student, GPS data, MIT, or home, that are used in defining access control policies. \textbf{Access control policies} are rules defined over a set of attributes using logical symbols, such as AND, OR, <, >, and =.  For example, a policy composed of four attributes, \textit{doctors}, \textit{trainers}, \textit{researchers}, and \textit{start\_date}, may be expressed as \textit{"doctor" OR "trainer" OR ("\textit{researcher}" AND "\textit{start\_date}" > Jan 1, 2023)}.  In CP-ABE, each piece of data is encrypted with its corresponding access control policy, while attributes associated with users are used to decrypt the data.
%\sq{I would argue Public Param is the KEK; Encryption is done with both public param and policy as input. } However, the latest implementation of NAC-ABE has dropped the idea of KEK as it's not an actual key but rather just a called access control policy. 
%The decryption key (DKEY) is associated with one or more attributes that users possess which determine 
%  \sq{This is key policy. Is this for KP-ABE instead? Please distinguish them. } 
%These attributes typically describe a user and are used to obtain the decryption key. Users possessing attributes that satisfy the policy can decrypt the content.
In the above example, in order to access data governed by the mentioned policy, a user must possess the attribute ``doctor'' or ``trainer'', or possess the ``researcher'' attribute while having a ``start\_date'' attribute after Jan 1, 2023.

\subsubsection{Overview} Below we explain the data naming and work flow of NAC-ABE-2020 (as depicted in Figure \ref{fig:nac-sequence-diagram}). 
We assume that all NDN entities involved in NAC-ABE are security bootstrapped~\cite{yu2021enabling}, and Attribute Authority is configured with initial access control policies. 

\paragraph{Key Setup} First, Attribute Authority (AA) generates \emph{public parameters} and a \emph{master key} (see Section \ref{sec:data-validation} for more details). The AA distributes the public parameters to the producers and consumers for data encryption and decryption, respectively. The public parameters are published as an NDN data packet. 
\begin{center}
%Public Params \\
%Interest Name = \name{/<access-manager-prefix>/PUBPARAMS} \\
Public Params Data Name = \name{/<AA-prefix>/PUBPARAMS/<version-number>}
\end{center}
In addition, the AA generates a decryption key (DKEY) for each legitimate consumer $C$ using the master key and $C$'s attributes (note that NAC-ABE-2020 uses CP-ABE so the DKEY is constructed using the consumer's attributes, not the consumer's access control policy).  The DKEY is shared with a consumer using the consumer's public key.  More specifically, the DKEY is encrypted using a symmetric key and the symmetric key is then encrypted using the consumer's public key. 
Both the encrypted DKEY and the encrypted symmetric key are packaged into a single NDN data packet with the following name: 
%Using a symmetric key for the encryption if DKEY helps avoids expensive public key encryption/decryption.
\begin{center}
%DKEY Interest Name = \name{/<AA-prefix>/DKEY/<consumer's identity block>} \\
%\sd {/AA-prefix/DKEY/<identity name block>/<signature>} old name --- consumer's was sending signed DKEY interest. AA does the signature verification. 
DKEY Data Name = \name{/<AA-prefix>/DKEY/<consumer-prefix>}
\end{center}
%Compromising the master key would result in a data breach of all the data encrypted using that ABE public parameter. The private key (DKEY), generated through the combination of public params, the master key, and the attributes (or the access control policy of users), creates a unique binding between them and ensures data access control. 
%The DKEY encryption serves two purposes: i) they can be distributed as regular NDN data packets\sq{It don't think it is a purpose of encryption; Encoding as Data is more like the step after encryption. }, and a consumer can fetch them using an Interest following the DKEY naming convention:
The consumer retrieves the encrypted DKEY using the above name and decrypts the DKEY using its private key. 
% \sq{It might be better to separate the setup process and data retrieval. }
%ii) only legitimate consumers will be able to decrypt the DKEY packet and gain access to it.

\paragraph{Data Encryption}   Each piece of data from a producer is encrypted using a symmetric Content Key (CK). The CK is further encrypted using the public parameters and an access control policy provided by the producer, which determines who can access this data. 
% It is a combination of one or more attributes, combined using logical symbols such as AND, OR, <, >, =\}\sq{It's the third time I am seeing this explanation of policy; Maybe you can make a paragraph to discuss about ABE with boolean formula policy. }. 
Once the encryption is complete, both the encrypted data and the encrypted CK are published as NDN data packets.
\begin{center}
%Content Key (CK)
%Interest Name = \name{/<producer's identity>/CK/<key-id>} \\
CK Data Name = \name{/<producer-prefix>/CK/<random-word>/ENC-BY/<access-policy>},
%key-id is the 32-bit secure random word.
\end{center}
where \name{<random-word>} is a 32-bit random word that uniquely identifies the CK, and \name{<access-policy>} is a string representation of the access control policy.

\paragraph{Data Decryption}  The consumer retrieves the encrypted data which contains the CK data name.  It fetches the CK data, decrypts the CK using the public parameters and its DKEY, and finally decrypts the data using the CK.

\subsubsection{Properties} \textbf{\emph{NAC-ABE offers more flexibility and scalability in supporting different access granularities than the original NAC}}.  For example, it can combine attributes in various ways using logic gates, such as \textit{AND, OR, <, >, =}, to generate many access control policies without incurring additional overhead in key generation.
%\sq{can be a lot more than n policies}, 
%by enabling precise definitions of attributes and policies for fine-grained access control over data. 
%With ABE policies, the system can delegate access to data at a significantly granular level without incurring any additional overhead in key generation. As a result, NAC-ABE scales very well as the number of consumers increases. 
%For example, if the system needs \textbf{\textit{n}} levels of granularities, it can combine attributes in various ways using logic gates such as \textit{AND, OR, <, >, =} to generate \textbf{\textit{n}} policies\sq{can be a lot more than n policies}, with each policy corresponding to a specific granularity.
\textbf{\emph{It also scales much better than NAC with the number of consumers}}. 
%On the consumer side, each consumer is assigned a set of attributes. 
For \textbf{\textit{m}} consumers, the AA needs to generate only \textbf{\textit{m}} DKEYs.  In contrast, the original NAC with \textbf{\textit{m}} consumers and \textbf{\textit{n}} access granularities requires \textbf{\textit{n}} KEKs, and, in the worst case, it needs to produce \textbf{\textit{m $\times$ n}} KDK data packets if all the consumers want to access data at all the granularities. 

\section{Use Case: Mobile Health}
\label{sec:use-case-mhealth}
%To explain the main objective of this paper and facilitate the discussion throughout the paper, let us consider an example scenario of a mHealth infrastructure. 
We have been developing a mobile health (mHealth) system over NDN \cite{dulal2022building}, which uses NAC-ABE to control access to the mHealth data.  In this section, we give an overview of the mHealth system, focusing on its security requirements and the issues we identified in NAC-ABE-2020.
% We have been developing a mobile health (mHealth) system over NDN, which uses NAC-ABE to control access to the mHealth data.  In this section, we give an overview of the mHealth system, focusing on its security requirements, along with issues within NAC-ABE-2020 which we discovered through the development process thus far.

Our mHealth system is comprised of the following entities: 
(1) \textbf{participants} who are involved in health studies or interested in monitoring their own health condition that use mobile devices (e.g., smartwatches and smartphones) to obtain sensor data related to their health and behavior;
(2) \textbf{data repositories} that store and serve the participants' mHealth data; 
(3) \textbf{healthcare providers}, such as doctors and nurses, who analyze the participants' data and relevant electronic health records to deliver personalized treatment plans; 
(4) \textbf{researchers} who study the mHealth data for various research purposes, e.g., to derive biomarkers, predict physiological and behavioral events, and assess the effectiveness and impact of mobile health technologies; 
(5) \textbf{other data users}, such as faculty, students, and personal trainers, who may use the data for education, fitness training, or other purposes; and 
(6) \textbf{mHealth policy administrators} who establish policies for the mHealth system to ensure compliance with data privacy laws (e.g., HIPAA, GDPR), patient consent requirements, and security protocols.

In our system, mHealth data is published under name prefixes that contain the organization, study name, participant ID, device, and data type in the name components. For example, suppose Alice is a participant in a study of diabetic patients conducted by mhealth.org, and her participant ID is id123, then Alice's heart rate data recorded by her smartwatch is published under \name{/org/mhealth/diabetes/id123/watch/heart-rate}, and the blood glucose data from her CGM sensor is published under \name{/org/mhealth/diabetes/id123/cgm/blood-glucose}.  The collected data may be further processed to derive contextual information to facilitate access control (see Section~\ref{sec:security-requirements}).

\subsection{Security Requirements}
\label{sec:security-requirements}

%The mHealth system requires only authorized entities produce and fetch the data. 
As mHealth data contains sensitive information of the participants, the system should ensure utmost security and privacy such that only authorized users can access the data as defined by their access policy. Moreover, the mHealth system needs to support \textbf{contextual access control}, i.e., data sharing based on the \textbf{context} of the data's production.
% \lz{instead of saying ``contextual access control'', this is about controlling access to the data produced under specific context}
The contextual information can include a semantic location (e.g., home, gym, and office), activity (e.g., walking, running, sleeping, eating, smoking, and non-smoking), time, and more. Such information is either in the raw data, e.g., time, or attached to the raw data after it is collected from the participants.  For example, semantic locations can be derived from GPS data that is collected alongside the raw sensor data, while activities can be inferred from accelerator and gyroscope data.

% Figure~\ref{fig:use-case-mhealth} shows an example of contextual access control.
% %Alice uses various mHealth devices, including a continuous glucose monitor (CGM) sensor, a smartwatch, and a smartphone, to collect different types of mHealth data such as blood sugar levels and heart rate in order to monitor her health. 
% Alice intends to share her blood glucose data with her doctor when she is at \textbf{\textit{home}},  her heart rate data generated at the \textbf{\textit{gym}} after \textbf{\textit{April 7, 2023}} with her trainer, and both her blood glucose data and heart rate data generated at \textbf{\textit{work}} with researchers.  In this scenario, the ``doctors", ``trainers", and ``researchers" represent the data consumers, while the ``gym", ``home", ``work'', and ``time" serve as the context.  

% \begin{figure}[tbp]
% 	\centering
% 	\includegraphics[scale=0.24]{figures/access-control-contex.png}
% 	\caption{Access Control based on Data Context in mHealth System}
% 	\label{fig:use-case-mhealth}
% \end{figure}

% In our system, the various context types and values are pre-defined.  However, new data users may join the system at any time, and their access rights are not from a pre-defined set.  They negotiate their access rights with the data owners, i.e., participants who generated the data, or the organization curating the data (if the participants have delegated the right to grant data access to the organization)  based on their needs. The access rights are then encoded in policies that specify what data they can access under what contexts.  

If we consider data names and contexts as attributes, then an access control policy can be expressed using these attributes, which makes NAC-ABE a natural solution to meet the above requirements.  
However, we encountered several issues when we tried to use NAC-ABE-2020 in our system (see Section~\ref{sec:identified-issues}). 

We also looked into several other works and found that they were either mostly based on CP-ABE or its variants (e.g., \cite{li2016attribute, li2014toward, da2015access}), focused on consumer attributes \cite{reddick2022aabac}, or were outdated with no implementation details (e.g., \cite{ion2013toward}). Therefore, they were not applicable to our use case.

\subsection{Identified Issues}
\label{sec:identified-issues}

\subsubsection{Mismatch between CP-ABE and Our Requirements} 
\label{requirement-mismatch}

NAC-ABE-2020 uses CP-ABE which requires the encryptor to know the access control policy for a piece of data when encrypting the data.  Typically, the attributes in a CP-ABE policy describe some properties of the users, e.g., organization, job, and age.  We could define policies using these user-based attributes for our system, but the types of users that may have the right to access a dataset are not fixed.  There may be users from new organizations with different job titles that are added to the system who need to access the dataset after the dataset is encrypted.  In this case, we need to change the policy to include the new user type and re-encrypt every CK with the new policy and re-publish the re-encrypted CKs, which incurs significant overhead for a large dataset as it may have many CKs.

The attributes in our access control policies describe properties of data, e.g., data names and contexts, but it is problematic if we use them in a CP-ABE approach, i.e., encrypting data with policies.  For example, we may encrypt a piece of data with the policy (\name{/org/mhealth/diabetes/id123/cgm/blood-glucose} AND ``home'') in order to limit the users to those who are authorized to access Alice's blood glucose data produced at home.  
Suppose a user is authorized to access two other types of data,  \emph{Alice's blood glucose data at work} and \emph{Alice's heart rate data at home}. This means the user owns four attributes -- \name{/org/mhealth/diabetes/id123/cgm/blood-glucose}, ``work'' , \name{/org/mhealth/diabetes/id123/watch/heart-rate}, as well as ``home'', which are encoded in their DKEY.  Since this user has both \name{/org/mhealth/diabetes/id123/cgm/blood-glucose}'' and ``home'' attributes, they incorrectly have access to Alice's blood glucose data produced at home.  Key-Policy Attribute-Based Encryption (KP-ABE)~\cite{goyal2006attribute} is a solution to this problem, as we will show in Section~\ref{sec:kp-abe-addition}.

\subsubsection{Lack of Trust Schema Support}
Trust schema \cite{yu2015schematizing} is an essential security component that systematically authenticates and authorizes packet signatures by leveraging semantic meaningful names.
Although the NAC-ABE library is capable of signing packets, the receiving end does not verify the legitimacy of the data using a trust schema.
The absence of systematic packet validation introduces potential risks of unauthorized data production by the attribute authority.

\subsubsection{Naming Issues and Inconsistencies}
\label{naming-issue}
The naming scheme in NAC-ABE-2020 does not strictly follow the original NAC-ABE design \cite{zhang2018nac}, which could lead to confusion and interoperability challenges. 
%For example, the PUBLIC PARAMS name does not include a component to distinguish between different ABE schemes,
In addition, the consumer's name prefix in the DKEY data name is insufficient for consumers to identify the appropriate private key to decrypt the DKEY, if consumers change their public keys but not their name prefixes.  The DKEY data name is also missing a version number to support any changes in the DKEY (e.g., when a consumer's attributes or policies change).

% \sd{The naming has been updated since the last paper, we might still need to talk about it somewhere}
% NAC-ABE \cite{zhang2018nac} follows NAC in naming all of its keys. 
% \textbf{Key Encryption Key (KEK)} is generated by attribute authority for each granularity as per the access policy. Encryptors are required to fetch the KEK to encrypt the symmetric content key CK. KEK is named as follows
% \begin{center}
% KEK Interest Name = /<access manager prefix>/NAC/<granularity>/KEK/
% KEK Data Name = /<access manager prefix>/NAC/<granularity>/KEK/<access-policy>	
% \end{center}
% Key Decrytion Key (KDK)
% Content Key (CK)
% \textbf{Naming Issues and Inconsistencies}

\subsubsection{High Overhead of Content Key Generation}
\label{content-key-generation}

NAC-ABE-2020 generates a new content key (CK) for every piece of data.  This approach can cause significant overhead for our mHealth system as it needs to support high-frequency sensor data.
 % While this approach may be suitable for applications that do not produce a large volume of data, it is not ideal for  
%\sq{I think another factor is that the size of mHealth data are relatively small, so having multiple Data sharing CK will not weaken the key. If there is an application that generates high-frequency of large data, they probably still need to have 1 CK per message. }
%Moreover, generating a new CK for every data point is \sq{overhead but it's not unnecessary}unnecessary overhead. 
The primary objective of generating new CKs is to mitigate potential damage in the event of a key compromise.  If the total size of the data protected by one CK is large, a compromised key could leak a substantial amount of data.  
On the other hand, if the CK generation is too frequent, the key generation and distribution may incur too much overhead and may not be necessary from a security viewpoint if the data size is small.  
% On the other hand, if the CK generation is more frequent, the key generation and distribution may incur too much overhead and may not provide a strong enough security improvement due to the smaller data size.
%, thus requiring a tradeoff to be considered.
Therefore, in order to achieve a good balance between security and overhead, the NAC-ABE library should support adjustable CK generation frequency based on data production rate and data size.

\subsubsection{Lack of Support for Large Key Sizes} 
\label{key-segmentation-issue}

NAC-ABE-2020 utilizes the OpenABE crypto library \cite{openabesite} for all cryptographic operations. 
% OpenABE represents attributes and policies as strings.
% When generating a DKEY using Key-Policy Attribute-Based Encryption (KP-ABE), as discussed in section \ref{kp-abe-addition}, the access policy or attribute set (e.g., attributes {``doctor", ``trainer"}, or policy {``doctor" AND ``trainer"}) needs to be passed as an argument to OpenABE. Furthermore, 
% Access policies given to OpenABE are logical expressions of entities of two types: (1) strings (e.g. ``doctor''); (2) attribute comparisons using operators such as $<$, $>$, $<=$, and $>=$.
Access policies given to OpenABE are logical expressions containing entities of two basic types: (1) strings (e.g. ``doctor'') and (2) attribute comparisons using operators such as $<$, $>$, $<=$, and $>=$ (e.g. ``floor > 2'').
Dates may also be used in these comparisons, which OpenABE internally converts into integer comparisons (day granularity) representing the specified dates or date ranges.
To facilitate integer comparisons, OpenABE converts n-bit integers (at most 32) into O(n) string attributes, leading to larger key sizes for policies with attribute comparisons.
% These attribute comparisons may be integers such as a room number, zip code, or UNIX timestamp, or a date or date range which are internally converted into two integer (UNIX timestamp) comparisons.
% An attribute within an attribute set or policy can be a string, date, or integer. 
% Date attribute comparisons, such as ``Date = Jan 01, 2025'', are internally converted into two integer (UNIX timestamp) comparisons to achieve a range from the beginning to end of the day.

% However, this was insufficient for our use case, as we wanted to facilitate access control down to second granularity. 
% To achieve this, our date/time granularity needed to be converted to a numerical timestamp before it could be used. For example, a policy like  (start\_date $>$ January 1, 2023, 10:00) needs to be converted to (start\_date $>$ 1672588800). 
% A 32-bit integer would result in approximately 64 attributes, 
% significantly increasing the overall size. 
% For example, an attribute ``number = 4 (100 in bits)", will be converted to three attributes num\_bit2=1 & num\_bit1=0 & num\_bit0=0 

To achieve access control at the second granularity, we use integer comparisons with 10-digit (32-bit) UNIX timestamps.
As a result, the size of a DKEY generated using time attributes (CP-ABE) or a policy containing time attributes (KP-ABE) is relatively large, and can easily surpass NDN's Maximum Segment Size (MSS).  
The CK data may also exceed the default segment size if the CK is encrypted with time attributes (CP-ABE) or a policy containing time attributes (KP-ABE).
For example, our measurements show that each comparison involving a time attribute increases the DKEY and CK sizes by about 6KB (Table \ref{tab:ck-dkey-size}).
Therefore, in order to support time-based attributes and policies, large keys need to be segmented when they are fetched, but this is not handled by NAC-ABE-2020. 
%\sq{How time and float convert to binary attribute is more of our speculation; maybe make the experimented size be the center of the paragraph. }

\begin{table}[]
    \centering
    \caption{Number of 32-bit Unix Timestamp (second granularity) comparisons in a policy vs DKEY Size (KP-ABE) and CK Size (CP-ABE)}
    \begin{tabular}{|p{1.8cm}|c|c|}
    %p{2cm}|p{1.3cm}|p{1.1cm}}
    \hline
    \# of integer-based attributes & DKEY Size (KB) & CK Size (KB)\\
    \hline
    1	& 6.96 &  7.03\\ \hline
    2	& 13.11 & 13.18\\ \hline
%    2	& 11.33 & 11.399\\ \hline
    3	& 19.06 & 19.13\\ \hline
    4	& 25.41 & 25.47\\ \hline
    5	& 31.56 & 31.62\\
    \hline
    \end{tabular}
    \label{tab:ck-dkey-size}
\end{table}
% \lz{I dont know what is the number of comparison, the first column in the table}
\section{Design and Implementation}
\label{design}
% \lz{design and implementation of what?}
% Overview of the new design
In this section, we describe the design and implementation changes we made to NAC-ABE-2020 in order to address the issues mentioned in the previous section.  
More specifically, we added KP-ABE (Key-Policy Attribute Based Encryption) which is a better fit than CP-ABE for our application (Section~\ref{sec:kp-abe-addition}), CK-caching producer (Section~\ref{sec:cache-producer}), data validation (Section~\ref{sec:data-validation}), and key segmentation (Section~\ref{sec:key-segmentation}).  
Additionally, we made changes to the NAC-ABE API and the message exchanges among NAC-ABE components (Section~\ref{sec:api-messages}), as well as the naming of the public params, DKEY, and CK data (Section~\ref{sec:naming}).   

Note that, although we made these changes for mHealth applications, the improved NAC-ABE can also provide better support to other applications.

\subsection{API and Message Exchanges}
\label{sec:api-messages}

\begin{figure*}[tbp]
	\centering
	\includegraphics[scale=0.45]{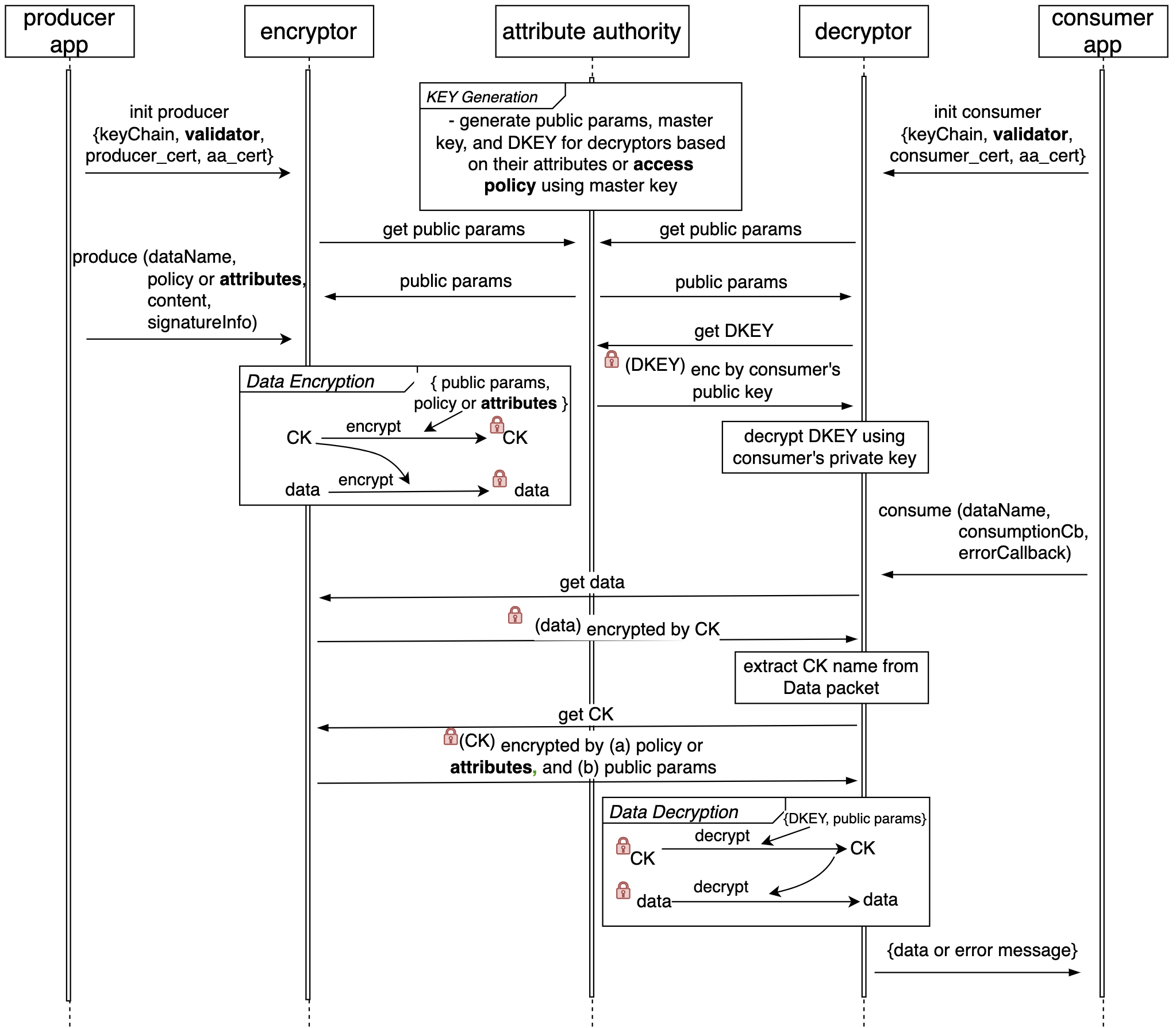}
	\caption{NAC-ABE Sequence Diagram (our changes are highlighted in bold.)}
	\label{fig:nac-sequence-diagram}
\end{figure*}

Figure~\ref{fig:nac-sequence-diagram} shows the interface between producer/consumer applications and NAC-ABE's encryptor/decryptor components, as well as the message exchanges among the encryptor, attribute authority, and decryptor within NAC-ABE.  Most of the API calls and messages remain the same as those in NAC-ABE-2020, as we described in Section~\ref{sec:nacabe-background}.  Nevertheless, we made modifications (highlighted in bold in Figure~\ref{fig:nac-sequence-diagram}) to support KP-ABE (Section~\ref{sec:kp-abe-addition}) and data validation (Section~\ref{sec:data-validation}).   For KP-ABE, we added (a) the option of using access control policy in the DKEY generation, and (b) the option of encrypting CK using attributes.  We also added a validator parameter to the API calls to support data validation.  We will describe these changes in more detail in the rest of this section.

\subsection{Naming}
\label{sec:naming}

\begin{figure}[tbp]
	\centering
	\includegraphics[scale=0.2]{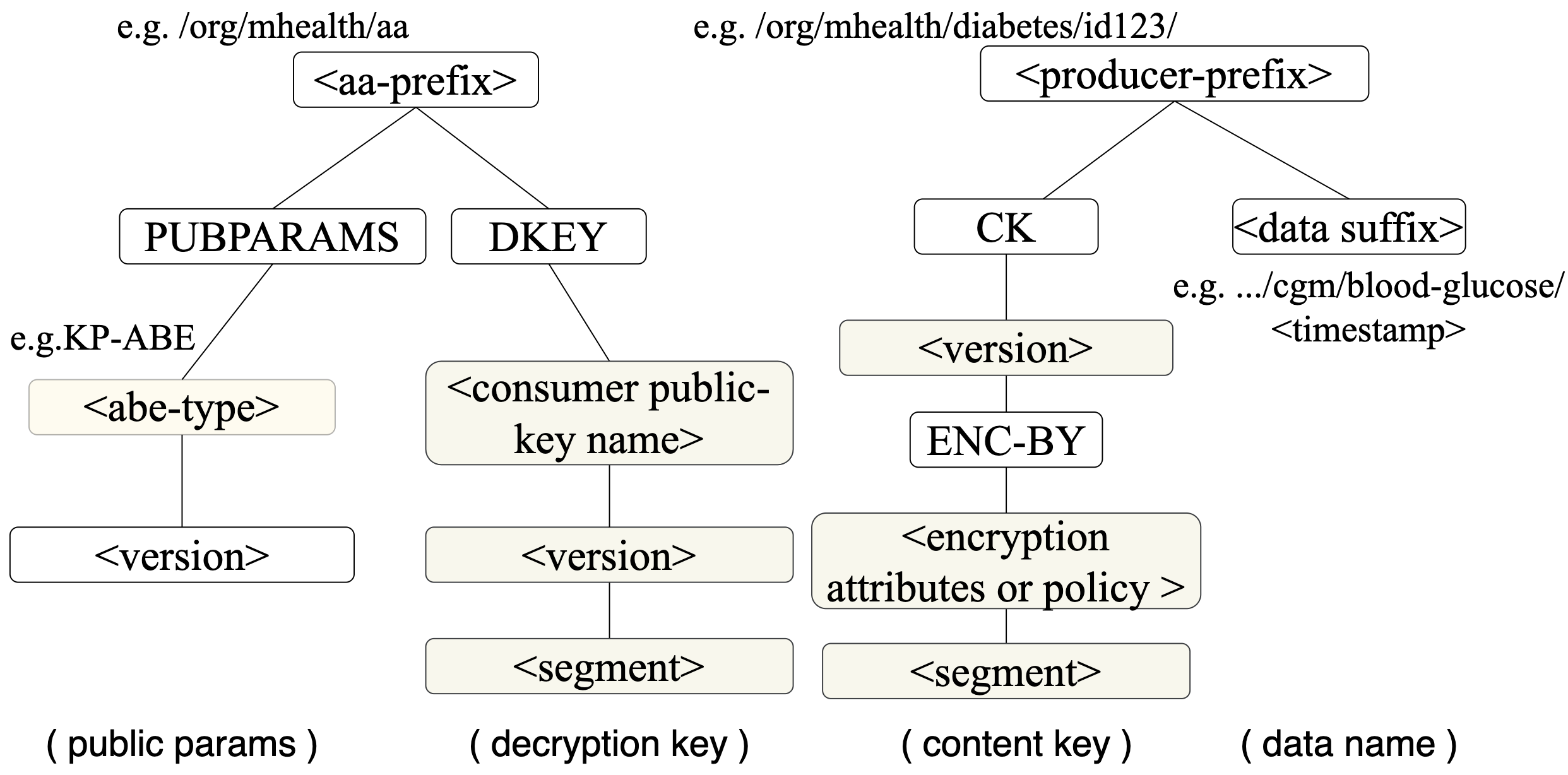}
	\caption{NAC-ABE Naming Scheme (our changes are in the highlighted boxes.)}
	\label{fig:data-key-naming}
\end{figure}

NAC-ABE publishes various keys and other data to facilitate authentication and access control. 
%These packets help different entities, such as producers, consumers, and the attribute authority, encrypt, sign, verify, and decrypt application data. 
We made several changes to the NAC-ABE-2020~\cite{zhang2020nacabe} naming scheme to support KP-ABE and key segmentation, as well as to address other naming issues. Figure \ref{fig:data-key-naming} shows the updated naming of different keys and data produced by NAC-ABE. 

\textbf{First}, we added a component \name{<abe-type>} to the public parameter naming to support both CP-ABE and KP-ABE. 
Following this change, an application can use one of the ABE types based on its needs. 
\textbf{Second}, since a consumer's public key can change over time, we replaced their name prefix in the DKEY naming with the name of their public key, making it easier for consumers to identify the appropriate private key for decrypting the DKEY.
\textbf{Third}, we added version and segment components to the DKEY naming. The version component helps keep track of the DKEY when it is changed, e.g., due to key rollover or attribute/policy changes. The segment component helps with key segmentation and reassembly whenever the size grows above the NDN segment size.
\textbf{Finally}, we made the following changes to the CK naming: (a) replaced the \name{<random-word>} with a version component to make it easier to track the CK; (b)  changed \name{<encryption policy>} to \name{<encryption attributes or policy>} to support KP-ABE; and (c) added a segment component to support CK data segmentation.

\subsection{KP-ABE Addition}
\label{sec:kp-abe-addition}

%\begin{figure}[htbp]
%	\centering
%	\includegraphics[scale=0.2]{figures/kp-abe-scheme.png}
%	\caption{Kp\sq{Complete the caption}}
%	\label{fig:contextual-access-control}
%\end{figure}

As mentioned in Section \ref{requirement-mismatch}, CP-ABE is not a suitable approach for our mHealth system because, in order to publish a piece of data, it needs to encrypt the data's CK with access control policies typically composed of user-based attributes, but the types of users who can access certain data may change over time.  
For example, initially only researchers from a particular lab in a University can access the data, but later, doctors from a Hospital  may also gain access to the data.  
% With CP-ABE, the CK needs to be re-encrypted and re-published every time the access control policy changes.  
With CP-ABE, the CK would need to be re-encrypted and re-published when this policy changes.
Attributes of the data such as name, location, and time, are generally known at the time of data publication and they do not change over time.  However, as we have shown in Section \ref{requirement-mismatch}, data-based attributes may not work well with CP-ABE.
% The attributes of the data, e.g., name, location, and time, are generally known by the producers at the time of data publication and they do not change over time.  However, data-based attributes may not work well with CP-ABE as we have shown in Section \ref{requirement-mismatch}.
%Furthermore, the access delegation can also change over time.

Key-Policy Attribute-Based Encryption (KP-ABE)~\cite{goyal2006attribute} offers a better solution to our problem. 
%Similar to CP-ABE, it is an attribute-based encryption scheme that provides fine-grained access control over encrypted data. 
In KP-ABE, producers are not required to specify the access control policy during encryption. 
Instead, they determine the data's attributes, which are then used to generate an encryption key (CK in our system) for that specific data.
The decryption key (DKEY) is associated with access control policies defined using the attributes. For example, Alice's device can encrypt her blood glucose data with the data name prefix \name{/org/mhealth/diabetes/id123/cgm/blood-glucose} and the location where the data was produced. Subsequently, she can work with attribute authority to delegate access rights to her doctor, trainer, and researchers, enabling them to obtain the right DKEY containing their access policies.  Thus, for our mHealth system, access delegation can be independent of the data publication, which means that changes in access control policies do not require re-encrypting the CKs. This provides better flexibility in terms of modifying access control policies.  

Below we briefly describe how our system uses KP-ABE based access control, as illustrated in Figure~\ref{fig:nac-sequence-diagram}.
\begin{itemize}
\item \emph{Setup:} the producer application determines a set of attribute types and values for encryption.  In our system, the attributes are data name, semantic location (home, work, gym, ...), and activities (e.g., sleeping, walking, driving, smoking, not smoking, ...).  
%For example, to encrypt blood-sugar data generated at home while smoking at a time on April 7, 2023, at 6 pm, Alice can use the attribute set {producer: Alice, datatype:blood-sugar, location: home, activity: smoking, timestamp: April 7, 2023, 6:00:00 pm}. Next, she has the option to define the access control policies for her data or collaborate with a trusted Attribute Authority (AA) to do so. For example, she can set a policy for her doctor as \name{Alice AND blood-sugar AND Home}. Similarly, she can define policies like \name{Alice AND heartRate AND gym AND Time > April 7, 2023, 6:00:00 pm} for her instructor, and so on.

\item \emph{Public Parameter and Master Key Generation:} the Attribute Authority (AA) is responsible for generating the public parameters and the master key.  The public parameters are published as an NDN data packet which is fetched by both the NAC-ABE encryptor and decryptor. 
\item \emph{DKEY Generation:} The AA also generates a DKEY for each consumer. These DKEYs are based on the consumer's individual access control policies and are encrypted using
%To make the DKEYs accessible to consumers, they are encrypted using a symmetric key, and the symmetric key is further encrypted using 
the consumer's public key.  The DKEY is fetched by the NAC-ABE decryptor.

\item \emph{Data Encryption:} the NAC-ABE encryptor encrypts each piece of data using a CK, and it encrypts the CK using the public parameters it fetched from the AA, along with a set of attributes supplied by the producer application for the specific data.
% The content key and the data are named as 
%\begin{center}
%    \name{ Data name: } \\
%    \name {Content key name: }
%\end{center}
%\sq{Remember to fill this in}
Subsequently, the name of the CK, encoded in TLV format, is encapsulated within the data packet. This ensures that the NAC-ABE decryptor can construct the name of the CK upon receiving the data packet.

\item \emph{Decryption:} 
%the consumer will first retrieve the public parameters and the DKEY from the AA. The DKEY packet contains a symmetric key that is decrypted using the consumer's private key. Afterward, the DKEY itself is decrypted using the symmetric key.
% \sq{Unless we agreed on an update, I don't think we had a symmetric key encryption; it's just private key wrapping DKEY}. 
% \sd {This is already implemented in the code}
the NAC-ABE decryptor extracts the CK name from the received data and retrieves the corresponding CK. The decryptor then uses the public parameters and the DKEY it fetched from the AA to decrypt the CK, and finally decrypts the data using the CK.  It then passes the data to the consumer application.
% This decryption process allows the consumer to access the content using the decrypted CK.
\end{itemize}

\subsection{CK-Caching Producer}
\label{sec:cache-producer}

The original NAC-ABE encryptor generates a new CK for every piece of data.  
We implemented a CK-Caching producer (on the basis of the original encryptor) that balances the overhead of generating new CKs with the security risks of data leaking due to compromised CKs. 
When a CK is generated for a specific policy or attribute set, it is cached for future use. Upon receiving new data for publication, the producer first checks if a CK already exists for the corresponding encryption attribute set or policy.  If so, that CK is used.  Otherwise, a new CK is generated.  In addition, to prevent using one CK for too much data, the CK is periodically updated based on configurable parameters set by the application. There are two conditions under which an update occurs: either after a predetermined number of data items have been encrypted using the CK, or a specific time interval has passed. These parameters can be adjusted according to the requirements of the application. Thus, with the CK-Caching Producer, we can reduce the overhead associated with generating CKs without compromising the security of the data. 

\subsection{Data Validation}
\label{sec:data-validation}

%Todo: add an example trust schema and explain it

We added data validation in NAC-ABE to check the authenticity of every data packet (including keys and application data) using a trust schema~\cite{yu2015schematizing} provided by the application.  
The trust schema specifies the trust hierarchy within the system, including a trust anchor and the relationship between data names and signing key names.  
In our system, the trust anchor is the organization that collects the mHealth data.   The trust anchor signs the public keys of the Attribute Authority, producers, and consumers.  The Attribute Authority signs the public parameters and DKEYs, which are validated by the NAC-ABE encryptors and decryptors.  Application data is signed by the NAC-ABE encryptor on behalf of its associated producer application, and validated by the NAC-ABE decryptor on behalf of its associated consumer application.

\begin{figure}[H]
	\centering
	\includegraphics[width=0.45\textwidth]{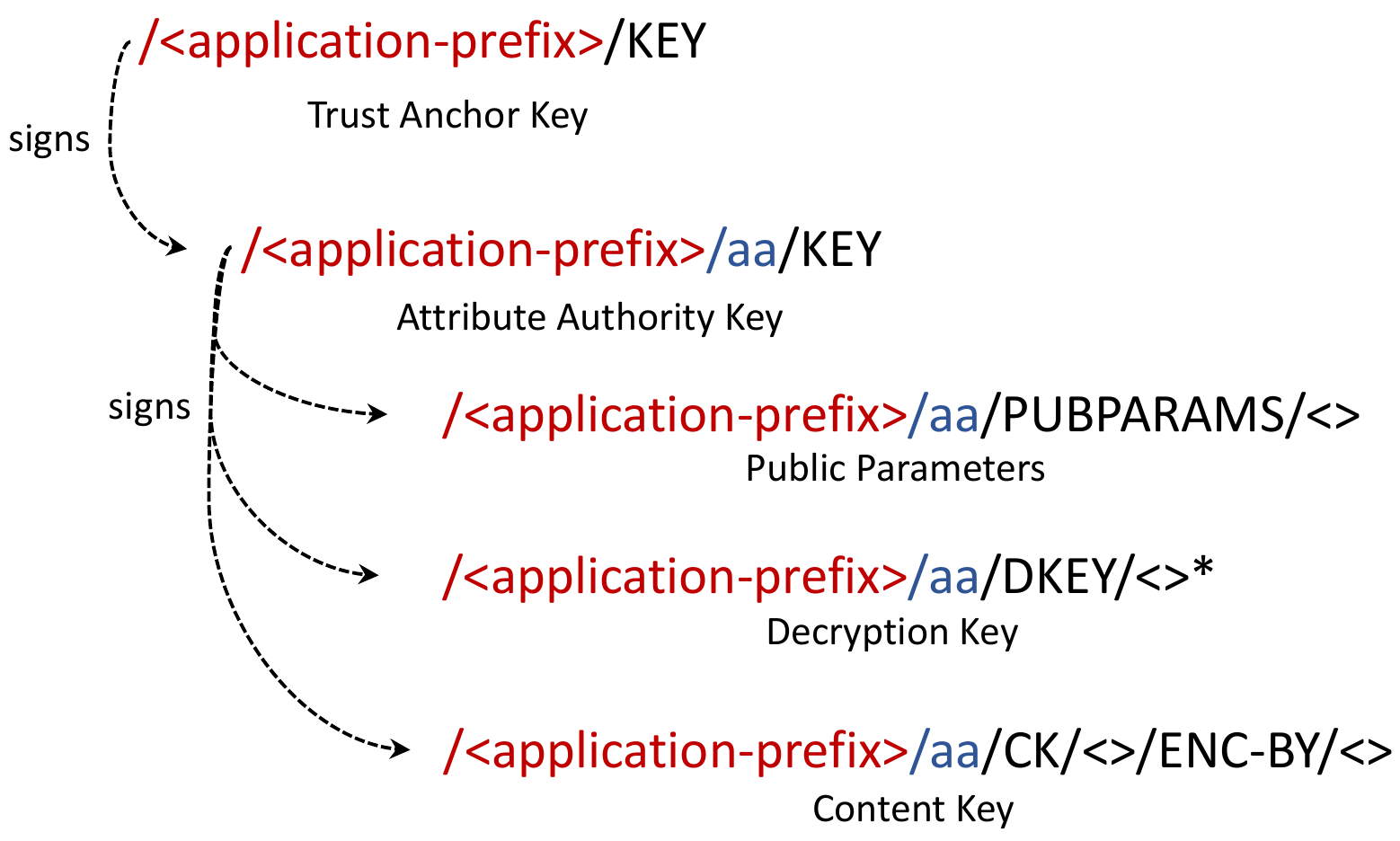}
	\caption{Application Trust Schema}
        \vspace{-0.5pt}
	\label{fig:trust-schema}
\end{figure}

Applications pass their trust schema to the NAC-ABE library at the application starting time.  NAC-ABE ensures that (i) keys and application data must follow the naming convention defined in Section~\ref{sec:naming}, and (ii) data must be signed by public keys that follow the hierarchy defined in the trust schema, and (iii) the data name prefix matches the data name prefix.   More specifically, upon receiving a Data packet, the NAC-ABE library checks if the signing key certificate in the Data KeyLocator is legitimate according to the trust schema.  If so, it expresses an Interest to fetch the certificate of the signing key, and cryptographically verifies the packet signature using the fetched public key.
The validator recursively validates every packet along the signing chain until reaching the trust anchor.

% Our trust schema support covers the Public Parameters, Content Keys, Decryption Keysm 

% -- Write about data validation and how it works in the current NAC-ABE design. \\
% -- validation of public params \\
% -- validation of the keys \\
% --- decryption keys \\
% --- application data \\
% --- content keys \\

\subsection{Key Segmentation}
\label{sec:key-segmentation}

We address the oversized key issue (Section~\ref{key-segmentation-issue}) by segmenting the DKEYs and CKs when they exceed the Maximum Segment Size (MSS).   Applications can pass the MSS to NAC-ABE.   Otherwise, NAC-ABE will choose the default value of 1500 bytes.  Each segment is an independently signed Data packet with the FinalBlockId field indicating the total number of segments.

When a decryptor fetches a key, it first expresses an Interest in the key name prefix.  Upon receiving the first segment, the decryptor learns the version number of the key and the total segment number from the FinalBlockId.
Afterward, the decryptor fetches all the segments sequentially.
During fetching, it uses a TCP-like slow-start with Additive Increase and Multiplicative Decrease (AIMD) as the congestion control algorithm.
The decryptor also validates each segment based on the trust schema.

\section{Discussion}
\label{sec:discussion-and-lessons-learned}
% \subsection{Discussions}
\noindent \textbf{Naming and Key Management:} We made several changes to the naming \ref{sec:naming} to incorporate KP-ABE, and key segmentation, and to address other issues \ref{naming-issue}. However, there are still several challenges still not resolved by our changes. For example, naming in the original NAC (Section \ref{sec:nac-background}) can achieve automatic key management. An encrypted data packet is enough for a consumer to fetch all the necessary keys to decrypt it. However, in NAC-ABE, the consumer can infer the CK name from an encrypted packet, but cannot construct either the PUBLIC PARAMS or DKEY name from the CK packet. The public parameters and DKEY are fetched independently, following the naming convention:

\begin{center}
    Public Params Interest: \name{/<aa-prefix>/PUBPARAMS} \\
    DKEY Interest: \name{/<aa-prefix>/DKEY/<consumers public key name>}
\end{center}

While this might be sufficient for the case when keys are never updated, this is not the case for us. In NAC-ABE, DKEY and Public Params can change, which is why we added a version number to the naming. This raises the question of how the consumer will fetch the keys with the correct version number. Also, the consumer can receive data packets encrypted with different public parameters and DKEYs. Therefore, the consumer needs to know not only the latest keys but also those that have been published in the past. Similarly, the producer needs to know the latest public parameters. Note that in some cases, the AA can generate keys, insert them into the repository, and then go offline. Therefore, sending an Interest with the MustBeFresh flag will not help either.

One possible solution to this issue is to rename the public parameters and DKEY to replace the version number with a key ID. Since the DKEY is generated using a specific master key and public parameters, they can share the same key ID. Additionally, the DKEY can change if the access policy changes, so the access policy or attribute set must also be included in the DKEY. The final naming would look as follows:

\begin{center}
    Public Params Interest: \name{/<aa-prefix>/PUBPARAMS/<abe-type>/<key-id>}
    
    DKEY Interest: \name{/<aa-prefix>/DKEY/<consumers public key name>/<access policy>/<key-id>/<segment>}
\end{center}

The content key is encrypted using both the public parameters and the policy (or attributes), so we can include the full name of the public parameters in the CK packet. Once the consumer receives the CK packet, it can extract the public parameters name and fetch the parameters. Additionally, assuming it knows its own access policy, it can construct the appropriate DKEY name using the key ID present in the public parameters name. Therefore, with these naming changes, we can achieve automatic key management on the consumer side. However, one challenge still remains: how will the producer know the latest public parameters?

% - \textbf{Discovering DKEY Version Number} \\
% -- Currently, we use the nan-cxx segment fetcher. It assumes that the last two components represent the version number. \\
% --- Is there a better solution? \\
% --- Are there potential scenarios where the current implementation won't work? provide an example? \\
% --- We could use RDR (Resource Discovery and Resolution) for version discovery and then the segment fetcher to retrieve the DKEY. \\

\noindent\textbf{Trust Schema for NAC-ABE:}
The issue of trust schema management in the NAC-ABE library presents a challenge that requires careful consideration.
On one hand, requiring the application to provide the trust schema adds a burden on developers, as they would need to be knowledgeable about NAC-ABE's trust models and protocol details (including the potential segmentation mentioned in Section~\ref{sec:key-segmentation}).
% On the other hand, embedding protocol-specific trust schema directly into the library is not straightforward either.
However, as discussed in Section~\ref{sec:key-segmentation}, data naming in NAC-ABE relies on context relating to the application namespace, which the library itself may not have.
The solution to this dilemma is currently under our investigation.

% the application? \\
% -- If the answer is "yes," application developers must have a deeper understanding of the protocol beyond just application logic. They need to understand the entire protocol, including details like segmentation because the layer beneath can have segmentation, so they must write rules that handle it. \\
% -- If the answer is "no," who should provide the trust schema? \\
% -- Why library can’t the write trust schema? \\
% --- The library may not have a complete context for writing a trust schema because applications use the protocol. \\
% --- The caller of the protocol should have the right to specify the trust policy. \\
% -- How can the gap between the application and the protocol library be bridged to define the trust schema? \\

% \textbf{- Lessons learned in policy design} \\
% -- Policies need to be scalable. Is the current policy integration scalable? If yes, how? If not, what needs to be done? \\

\noindent \textbf{Alternative crypto library: } NAC-ABE relies on the OpenABE crypto library, but lacks active maintenance since 2019 and is incompatible with newer versions of OpenSSL (v3). This issue can lead to deployment failures on future machines and systems, highlighting the need for an alternative crypto library. 
In addition to deployment issues and key length problems we discussed in section \ref{key-segmentation-issue}, OpenABE does not have a NOT operator, leaving application designers to either not facilitate a policy that denies access to data with certain attributes, or implement their own application-specific ways to have that functionality.

Several open-source options like Charm, GoFE, and Rabe have been suggested by Mosteiro-Sanchez et al \cite{mosteiro2022too}. However, we discovered that these libraries are also not actively maintained, face challenges with adapting to updates, and have unresolved issues. 
While popular libraries may have more reported issues, the overall maintenance and stability of these options remain uncertain. 
Therefore, finding a stable and adaptable crypto library for NAC-ABE is still an open issue.

\section{Conclusion and Future Work}
\label{conclusion-future-work} 
% \vspace{-1pt}
In this paper, we addressed the challenges and limitations of the existing NAC-ABE library and discussed the enhancements we made to facilitate fine-grained access control in data sharing, specifically for mHealth applications. We improved the security, flexibility, and efficiency of NAC-ABE by incorporating KP-ABE, key segmentation, and data validation, and by introducing changes to the naming scheme. Furthermore, we introduced a CK-Caching producer for optimized CK updates and addressed the issue of handling large DKEYs and CKs. Through our contributions, we provide a more robust ABE framework that can be used to enforce access control in NDN-based systems. The identified issues and potential future directions outlined in this paper offer valuable insights that can be used to further advance NAC-ABE libraries.

% \bibliographystyle{IEEEtran}
% \bibliography{refs}

\bibliographystyle{acm}
\bibliography{refs}

\begin{thebibliography}{10}

\bibitem{belguith2018phoabe}
{\sc Belguith, S., Kaaniche, N., Laurent, M., Jemai, A., and Attia, R.}
\newblock Phoabe: Securely outsourcing multi-authority attribute based
  encryption with policy hidden for cloud assisted iot.
\newblock {\em Computer Networks 133\/} (2018), 141--156.

\bibitem{bethencourt2007ciphertext}
{\sc Bethencourt, J., Sahai, A., and Waters, B.}
\newblock Ciphertext-policy attribute-based encryption.
\newblock In {\em 2007 IEEE symposium on security and privacy (SP'07)\/}
  (2007), IEEE, pp.~321--334.

\bibitem{da2015access}
{\sc Da~Silva, R.~S., and Zorzo, S.~D.}
\newblock An access control mechanism to ensure privacy in named data
  networking using attribute-based encryption with immediate revocation of
  privileges.
\newblock In {\em 2015 12th Annual IEEE Consumer Communications and Networking
  Conference (CCNC)\/} (2015), IEEE, pp.~128--133.

\bibitem{diffie2022new}
{\sc Diffie, W., and Hellman, M.~E.}
\newblock New directions in cryptography.
\newblock In {\em Democratizing Cryptography: The Work of Whitfield Diffie and
  Martin Hellman}. 2022, pp.~365--390.

\bibitem{dulal2022building}
{\sc Dulal, S., Ali, N., Thieme, A.~R., Yu, T., Liu, S., Regmi, S., Zhang, L.,
  and Wang, L.}
\newblock Building a secure mhealth data sharing infrastructure over ndn.
\newblock In {\em Proceedings of the 9th ACM Conference on Information-Centric
  Networking\/} (2022), pp.~114--124.

\bibitem{goyal2006attribute}
{\sc Goyal, V., Pandey, O., Sahai, A., and Waters, B.}
\newblock Attribute-based encryption for fine-grained access control of
  encrypted data.
\newblock In {\em Proceedings of the 13th ACM conference on Computer and
  communications security\/} (2006), pp.~89--98.

\bibitem{ion2013toward}
{\sc Ion, M., Zhang, J., and Schooler, E.~M.}
\newblock Toward content-centric privacy in icn: Attribute-based encryption and
  routing.
\newblock In {\em Proceedings of the 3rd ACM SIGCOMM workshop on
  Information-centric networking\/} (2013), pp.~39--40.

\bibitem{li2016attribute}
{\sc Li, B., Huang, D., Wang, Z., and Zhu, Y.}
\newblock Attribute-based access control for icn naming scheme.
\newblock {\em IEEE Transactions on Dependable and Secure Computing 15}, 2
  (2016), 194--206.

\bibitem{li2014toward}
{\sc Li, B., Wang, Z., Huang, D., and Zhu, Y.}
\newblock Toward privacy-preserving content access control for information
  centric networking.
\newblock Tech. rep., ARIZONA STATE UNIV TEMPE OFFICE OF RESEARCH AND SPONSORED
  PROJECT ADMINISTRATION, 2014.

\bibitem{mosteiro2022too}
{\sc Mosteiro-Sanchez, A., Barcelo, M., Astorga, J., and Urbieta, A.}
\newblock Too many options: A survey of abe libraries for developers.
\newblock {\em arXiv preprint arXiv:2209.12742\/} (2022).

\bibitem{NDN-TR1}
{\sc {NDN project team}}.
\newblock {Named Data Networking (NDN) Project}.
\newblock Technical Report NDN-0001, NDN, Oct. 2010.

\bibitem{openabesite}
{OpenABE Library GitHub Site}.
\newblock \url{https://github.com/zeutro/openabe}.

\bibitem{reddick2022aabac}
{\sc Reddick, D., Presley, J., Feltus, F.~A., and Shannigrahi, S.}
\newblock Aabac--automated attribute based access control for genomics data.
\newblock {\em arXiv preprint arXiv:2204.04591\/} (2022).

\bibitem{sahai2005fuzzy}
{\sc Sahai, A., and Waters, B.}
\newblock Fuzzy identity-based encryption.
\newblock In {\em Advances in Cryptology--EUROCRYPT 2005: 24th Annual
  International Conference on the Theory and Applications of Cryptographic
  Techniques, Aarhus, Denmark, May 22-26, 2005. Proceedings 24\/} (2005),
  Springer, pp.~457--473.

\bibitem{sim2019mobile}
{\sc Sim, I.}
\newblock Mobile devices and health.
\newblock {\em New England Journal of Medicine 381}, 10 (2019), 956--968.

\bibitem{yu2021enabling}
{\sc Yu, T., Moll, P., Zhang, Z., Afanasyev, A., and Zhang, L.}
\newblock Enabling plug-n-play in named data networking.
\newblock In {\em MILCOM 2021-2021 IEEE Military Communications Conference
  (MILCOM)\/} (2021), IEEE, pp.~562--569.

\bibitem{yu2015schematizing}
{\sc Yu, Y., Afanasyev, A., Clark, D., Claffy, K., Jacobson, V., and Zhang, L.}
\newblock Schematizing trust in named data networking.
\newblock In {\em proceedings of the 2nd ACM Conference on Information-Centric
  Networking\/} (2015), pp.~177--186.

\bibitem{yu2015name}
{\sc Yu, Y., Afanasyev, A., and Zhang, L.}
\newblock Name-based access control.
\newblock {\em Named Data Networking Project, Technical Report NDN-0034\/}
  (2015).

\bibitem{zhang2014named}
{\sc Zhang, L., Afanasyev, A., Burke, J., Jacobson, V., Claffy, K., Crowley,
  P., Papadopoulos, C., Wang, L., and Zhang, B.}
\newblock Named data networking.
\newblock {\em ACM SIGCOMM Computer Communication Review 44}, 3 (2014), 66--73.

\bibitem{zhang2020nacabe}
{\sc Zhang, Y., Zhang, Z., and Tu, Y.}
\newblock {CP-ABE} based {NAC-ABE} implementation, 2020.

\bibitem{zhang2018nac}
{\sc Zhang, Z., Yu, Y., Ramani, S.~K., Afanasyev, A., and Zhang, L.}
\newblock {NAC}: Automating access control via {Named Data}.
\newblock In {\em IEEE Military Communications Conference (MILCOM)\/} (2018),
  pp.~626--633.

\bibitem{zhang2018overview}
{\sc Zhang, Z., Yu, Y., Zhang, H., Newberry, E., Mastorakis, S., Li, Y.,
  Afanasyev, A., and Zhang, L.}
\newblock An overview of security support in named data networking.
\newblock {\em IEEE Communications Magazine 56}, 11 (2018), 62--68.

\end{thebibliography}

\end{document}